\documentclass[10pt]{iopart}
\usepackage{graphicx}% Include figure files
\usepackage{bm}% bold math
\usepackage{array}
\usepackage{cite}
\usepackage{setspace}
\usepackage{xcolor}
\usepackage{iopams}
\begin{document}

\title[]{Ytterbium lattice clock with systematic uncertainty of $1.3\times 10^{-18}$ and instability at the $10^{-19}$ level}

\author{Qiang Zhu$^{1}$, Jingran Shi$^{1,2}$,  Yuechen Zhang$^{1,2}$, Xiatian Xu$^{1,2}$, Bing Wang$^{1}$, Dezhi Xiong$^{1,3}$, Zhuanxian Xiong$^{1,3}$, Pengcheng Fang$^{1}$, Qunfeng Chen$^{1}$, Lingxiang He$^{1,3}$, and Baolong Lyu$^{1,3}$}

\address{$^{1}$ Innovation Academy for Precision Measurement Science and Technology, Chinese Academy of Sciences, Wuhan 430071, China}
\address{$^{2}$ University of Chinese Academy of Sciences, Beijing 100049, China}
\address{$^{3}$ Hefei National Laboratory, Hefei 230088, China}
\ead{zxxiong@apm.ac.cn}
\vspace{10pt}

\begin{indented}
\item[]\today
\end{indented}

\begin{abstract}
We report an optical lattice clock based on $^{171}$Yb atoms with a total systematic uncertainty of $1.3\times 10^{-18}$. An in-vacuum buildup cavity was employed to enhance the lattice light power. Differential frequency measurement between two identical clocks facilitates the evaluation of systematic shifts. Synchronous comparison of the two clocks reached a stability level of $2.7\times 10^{-19}$ in an averaging time of 216,000~s. The blackbody radiation (BBR) shield which is placed in vacuum provides a well-characterized BBR environment, enabling an uncertainty contribution of $8.7\times 10^{-19}$ from the BBR Stark shift. Lattice light shifts were measured at different lattice depths $U$, and fitted with a fourth-order polynomial of $U$. The zero-linear-shift frequency $\nu_{\mathrm{zero}}$ was determined to be 394 798 260.6(3) MHz. The lattice light shift can be controlled at an uncertainty level of $6.3\times 10^{-19}$ under typical operating conditions. Other systematic shifts have also been evaluated. The two clocks will be used for remote frequency comparisons between Shanghai and Wuhan.
\end{abstract}
\vspace{2pc}

\submitto{\MET}
\maketitle
\ioptwocol

\section{Introduction}\label{Introduction}

Optical clocks have advanced rapidly in the recent decade, with fractional systematic uncertainty reaching the $10^{-18}$ level \cite{Ushijima(RIKEN)2015,Huntemann(PTB)2016,McGrew(NIST)2018,Ohmae(RIKEN)2021,Huang(APM)2022,Cui(APM)2022,Tofful(NPL)2024,Ma(HUST)2024,
Liao(NIM)2025,Lu(NTSC)2025,Zhang(ECNU)2026} and beyond~\cite{Brewer(NIST)2019,Zhang(NUS)2023,Aeppli(JILA)2024,Marshall(NIST)2025,
Lindvall(Finland)2025,Zhang(APM)2026,Jia(USTC)2026}.  In the updated version of the roadmap towards the redefinition of the SI second, fractional uncertainty $\lesssim 2\times 10^{-18}$ is identified as a milestone criterion that an optical clock needs to reach~\cite{Dimarcq(Roadmap)2024}.
With increasing robustness, optical clocks based on a variety of atomic species are able to steer the time scales~\cite{Ido2016,Grebing(PTB)2016,Hachisu(NICT)2018,Yao(NIST)2018-2,Kobayashi(AIST)2020,Zhu(NIM)2024,Yuan(APMtimescale)2026}.
Particularly, recently developed transportable Sr clocks have achieved performances comparable to the best laboratory-based systems ~\cite{Ohmae(RIKEN)2021,Nosske(PTB)2025}. Compared to single-ion optical clocks, optical lattice clocks (OLCs) operating with a large number of neutral atoms significantly reduce the quantum projection noise, enabling higher clock stability. Long-term stabilities at $10^{-19}$ level have been reached in Sr and Yb lattice clocks \cite{Oelker(JILA)2019,McGrew(NIST)2018,Liu(USTC)2025}.

The systematic uncertainty of an OLC is mainly limited by the blackbody radiation (BBR) Stark shift and lattice light shift. For a lattice clock operated at room temperature, the BBR environment around the atoms needs to be well controlled and characterized~\cite{Beloy(NIST)2014,Xu(ECNU)2016,Xiong(APM)2021,Heo(KRISS)2022,Yu(USTC)2026}.
Alternatively, OLCs can operate in cryogenic environments to suppress the BBR shift, achieving a BBR shift uncertainty on the order of $10^{-19}$ \cite{Ushijima(RIKEN)2015,Nosske(PTB)2025} or even down to the low $10^{-20}$ level~\cite{Hassan(NISTcryo)2025}.

Lattice light shift evaluation requires determination of both the magic frequency and the atomic parameters associated with multipolarizabilities and hyperpolarizability. To precisely measure the hyperpolarizability shift, it is beneficial to push the lattice trap depth to a level which is more than 20 times the normal operational depth~\cite{Brown(NIST)2017}. A buildup cavity is usually used to enhance the lattice light power while a large beam size enables small density-dependent collisional shifts.

Here, we present our new-generation ytterbium OLCs (Yb1 and Yb2). The two clocks share the same design. Synchronous and anti-synchronized comparisons of them demonstrate fractional instabilities of $2.7\times 10^{-19}$ and $5.1\times 10^{-19}$, respectively, at the averaging time of 216,000~s. The full systematic uncertainty budget presented in this work is only for Yb2, and a complete evaluation of Yb1 is ongoing and expected to be reported elsewhere.

Compared to the previous generation \cite{ZhangAng(APM)2022}, the major design change to the vacuum system is the installation of an in-vacuum buildup cavity which enhances the lattice light power by two orders of magnitude. The larger beam waist significantly reduces the collision shift of the lattice-trapped atoms. More importantly, the enhanced optical lattice allows exaggerated light shift, and hence more precise determination of the related fit parameters. The uncertainty of the lattice light shift is reduced to $6.3\times 10^{-19}$ under normal operating conditions. To lower the frequency drift of the clock laser,  the previous ultra-low expansion (ULE) cavity used for frequency locking is replaced by another 30-cm-long ULE cavity (Stable Laser Systems), which operates at $1156~\mathrm{nm}$ and has an easily accessible zero expansion temperature (\mbox{19.4 $^{\circ}$C}). In-vacuum BBR shield with similar design to that from NIST~\cite{Beloy(NIST)2014} provides a homogeneous thermal radiation environment to the enclosed atoms, enabling a BBR Stark shift uncertainty below $10^{-18}$. The total systematic shift uncertainty is below $2\times 10^{-18}$, meeting the first threshold required for redefinition of the second.

\section{Setup and spectroscopy}\label{sec:Setup}
%Here, $E_{r}=h^{2}/\left ( 2m \lambda_{lat}^{2}\right )$ is the recoil energy of the lattice photon, where $h$ is Plank's constant, $m$ is the %mass of $^{171}$Yb and $\lambda_{lat}=759\,\mathrm{nm}$ is the wavelengh of the optical lattice.
\begin{figure*}[htp]
\centering
\includegraphics[width=0.85\textwidth,angle=0]{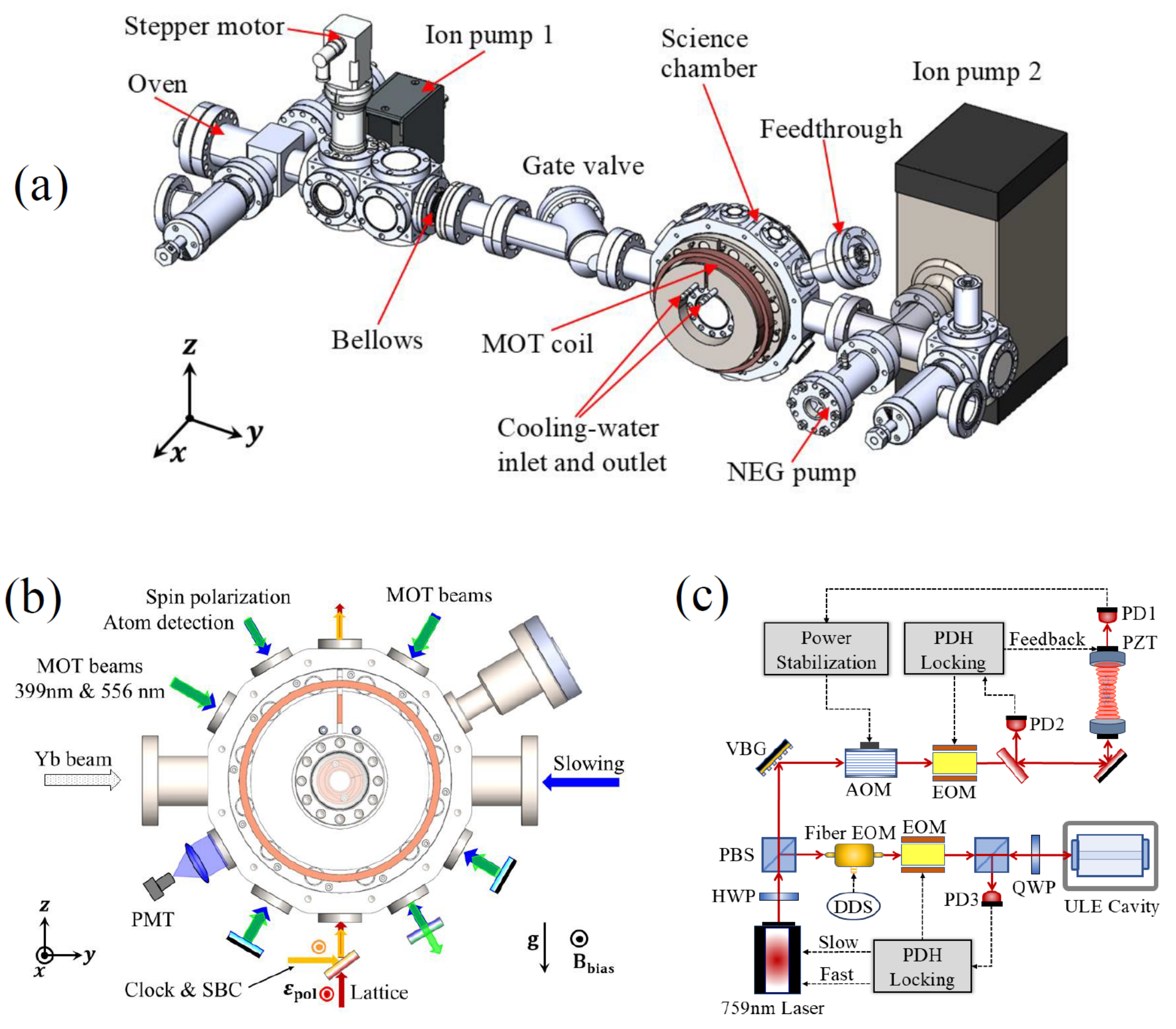}
\caption{Physics package and optical scheme of the Yb2 lattice clock. (a) Three-dimensional drawing of the vacuum system. The main components are marked. A BBR shield (not shown) is installed inside the science chamber. (b) Optical scheme for the science chamber. The directions of magnetic field, gravity orientation, and polarizations of the clock and lattice lasers are indicated. (c) Schematic of the vertically oriented lattice buildup cavity with in-vacuum mirrors. The cavity is locked to the 759~nm lattice laser. Photodiode (PD), polarizing beam splitter (PBS), acousto-optical modulator (AOM), electro-optic modulator (EOM), piezoelectric transducer (PZT), half-wave plate (HWP), quarter-wave plate (QWP), volume Bragg grating (VBG), direct digital synthesiser (DDS).}
\label{Fig:Setup}
\end{figure*}
As shown in figure\,\ref{Fig:Setup}(a), the vacuum system of Yb2 consists mainly of an atomic oven and a science chamber. Yb1 has an identical design in physics package except that the ion pump in the low vacuum region of Yb1 has a higher pumping speed. A BBR shield is installed inside
the science chamber to create a thermally homogenous environment for the enclosed $^{171}$Yb atoms. An atomic shutter blocks the atomic beam during spectroscopy, preventing the excess thermal radiation from the hot oven as well as collisions from the atomic beam. This shutter is actually a titanium bar with bored holes on side, which is fixed on a magnetically coupled rotary motion feedthrough (Kurt J. Lesker; MD40NDX000Z). Rotary motion driven by a stepper motor generates only weak mechanical vibrations. The vacuum system of each clock is placed on a separate breadboard, while the two breadboards for Yb1 and Yb2 reside on the same experimental table.

The most important improvement over the previous generation of clock is the buildup cavity for optical lattice, which is formed with two in-vacuum mirrors mounted on the BBR shield (see figure\,\ref{Fig:BBR-shield}(a)). The 759~nm lattice light is derived from a commercial laser (Precilasers; FL-SF-759-1.2-CW) which is based on frequency-difference generation between two fiber lasers. While the lattice laser is locked on a ULE cavity as shown in figure\,\ref{Fig:Setup}(c), the vertically oriented cavity is locked to the lattice laser, achieving a power enhancement factor of $\sim$150. At an input power of 150~mW, the cavity enhanced lattice reaches a depth of $1000\,E_{r}$. The large beam radius of $\sim\,165\,\mu\mathrm{m}$ is favorable for reducing the atomic density and hence the collisional shift. Additionally, in order to suppress the excessive atomic heating caused by lattice intensity fluctuations, the intra-cavity power is stabilized by a servo loop. As shown in figure\,\ref{Fig:Setup}(c), the transmitted lattice light power is detected by PD1. The error signal, obtained by subtracting the detected power from the set value, is fed to a proportional-plus-double-integral ($\mathrm{PI}^2$) controller which adjusts the RF power of the AOM to stabilize the intra-cavity power.
\begin{figure}[htp]
\centering
\includegraphics[width=0.99\columnwidth,angle=0]{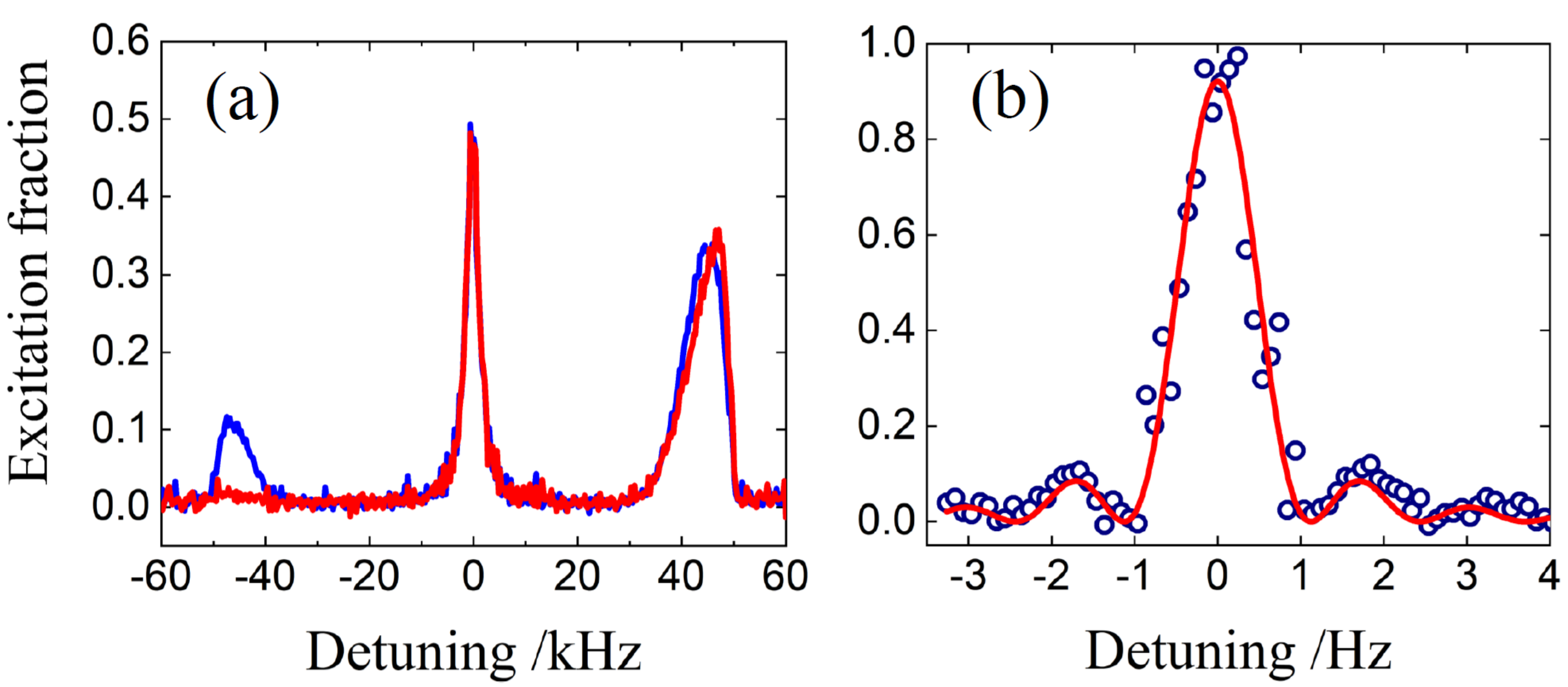}
\caption{(a) Longitudinal sideband spectra at $173\,E_{r}$ before (blue line) and after (red line) sideband cooling. The 16-ms sideband cooling reduces the longitudinal temperature from $1.9\,\mu\mathrm{K}$ to $0.8\,\mu\mathrm{K}$, corresponding to a mean longitudinal motional number $\langle n_{z}\rangle =0.04$. (b) A typical single-scan, 800-ms Rabi spectrum of the $^{1}$S$_{0}$-$^{3}$P$_{0}$ clock transition; the red line is a free-parameter sinc$^{2}$ function fit.}
\label{Fig:Spectroscopy}
\end{figure}

The atomic beam is decelerated by a laser beam red-detuned from the $^{1}$S$_{0}$-$^{1}$P$_{1}$ transition at 399~nm. Preparation of cold atoms, as well as the subsequent Rabi spectroscopy, are all performed inside the science chamber. In figure\,\ref{Fig:Setup}(b) we present the optical scheme around the science chamber.  Not shown is the 1388~nm laser beam which is used to optically pump nearly all the atoms in the $^{3}$P$_{0}$ state back to the ground state through the $^{3}$D$_{1}$ state. The operations that must be implemented in a typical clock cycle, including the two-stage magneto-optical trap (MOT), lattice loading, spin polarization, sideband cooling, Rabi interrogation, and atom detection, have been described in detail previously~\cite{ZhangAng(APM)2022}. Spin polarization and  sideband cooling are actually applied to the atoms at the same time, lasting for 16~ms. Sideband cooling pumps the atoms in the excited longitudinal motional states into the ground state of $n_{z}=0$, leading to a suppressed red sideband of the trapped atoms, as shown in figure~\ref{Fig:Spectroscopy}(a). The average motional quantum number $\langle n_{z}\rangle$ is about 0.04 after sideband cooling.

The clock laser locked to the ULE cavity is shared by both Yb systems. The $^{1}$S$_{0}$-$^{3}$P$_{0}$ clock transition of the spin-polarized atomic sample is interrogated by a $\pi$-polarized probe light pulse. An external magnetic field of $1~\mathrm{G}$ is applied to split the $\pi$ transitions with opposite nuclear spins ($m_{F}=\pm 1/2$). At normal operating conditions, the lattice depth is set to a moderate level around $50\,E_{r}$. The Rabi spectrum shows a Fourier limited linewidth for an interrogation time up to 800~ms (see figure~\ref{Fig:Spectroscopy}(b)).
\begin{figure}[htp]
\centering
\includegraphics[width=0.99\columnwidth,angle=0]{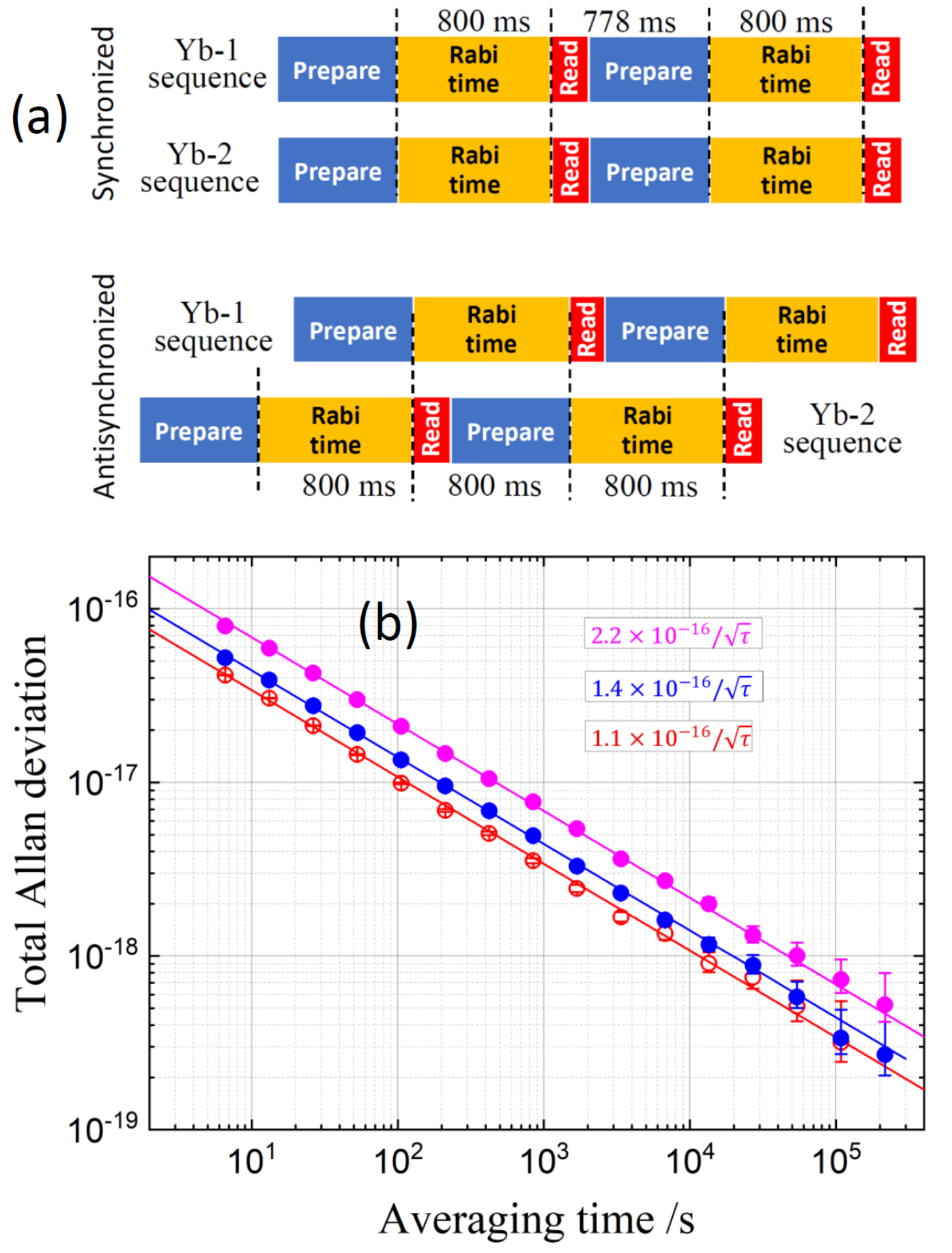}
\caption{(a) Time sequence for synchronized and anti-synchronized Rabi spectroscopy. (b) Measured instability of a Yb OLC. Total Allan deviation represents the single-clock measurement instability, $1/\sqrt{2}\left(\nu_{2}-\nu_{1}\right)/\nu_{\mathrm{clock}}$. The blue and red datasets use synchronized Rabi spectroscopy. The magenta set uses
anti-synchronized Rabi spectroscopy. The lines represent the white frequency noise asymptotes. For red dataset, Yb1 and Yb2 share the same 759~nm lattice laser. For other datasets, each clock uses its own lattice laser. Error bars represent 1$\sigma$ uncertainty.}
\label{Fig:FreqStability}
\end{figure}

The atomic preparation procedure is as follows. First, the optical lattice is loaded at a deep level ($1040\,E_{r}$). The deep lattice is chosen to provide a high loading efficiency. This is followed by a ramp-down to $52\,E_{r}$ in 50~ms. Then, spin polarization and sideband cooling are applied to the trapped atoms. After that, the optical lattice is adiabatically ramped to the desired level in 100~ms.

During normal clock operation, we alternately probe the half maximum points of the $\pi$-transition from each $m_{F}$ state. We take average of the two resonance frequencies to cancel the first-order Zeeman shift. The second-order Zeeman shift is corrected in real time, and its calibration method will be described in Section~\ref{subsec:OtherSystematics}. The BBR shift is updated and corrected every 10~s according to the readings from the thermal sensors on the BBR shield.

The frequency stability of an OLC is usually limited by Dick effect. However, the Dick noise can be common-mode rejected in the measured combined clock instability of two clocks if the atomic systems are interrogated synchronously. In the configuration of synchronized 800-ms Rabi spectroscopy, frequency comparison between Yb1 and Yb2 demonstrated a single clock instability of $1.1\times 10^{-16}/\sqrt{\tau}$, with $\tau$ the averaging time in seconds, reaching low $10^{-19}$ at an averaging time of $10^{5}$~s (see the red line in figure~\ref{Fig:FreqStability}(b)).  The frequency of the lattice laser shared by the two clocks was set to $394,798,266.9$~MHz, which is the magic frequency reported by the NIST group~\cite{Bothwell(NIST)2025}.

We recently received a new lattice laser of the same model, allowing each clock to operate with an independent lattice laser. With both lattice lasers held close to the magic frequency, we repeated the instability measurement for synchronized Rabi spectroscopy. Clocks Yb1 and Yb2 were operated at a depth level of $48\,E_{r}$ and $52\,E_{r}$, respectively. Lattice-trapped atomic numbers of Yb1 and Yb2 were $17,000$ and $24,000$, respectively. As shown in figure~\ref{Fig:FreqStability}(b), a single clock instability of $1.4\times 10^{-16}/\sqrt{\tau}$ was achieved, reaching $2.7\times 10^{-19}$ at $216,000$~s as determined by the last point of the total Allan deviation. In this measurement, the dataset was taken over a course exceeding 7 days, with an uptime of 96.8\%. We also measured the frequency instability in the configuration of anti-synchronized comparison. Optical lattice depths were unchanged, while the atom numbers were approximately equal ($11,000$ for Yb1 and $12,000$ for Yb2). The time sequence of clock cycles is shown in figure~\ref{Fig:FreqStability}(a). This dataset corresponds to a measurement of 9 days, with an uptime of 94.8\%. Since the Dick noise is maximized, the frequency instability was degraded to $2.2\times 10^{-16}/\sqrt{\tau}$. Nevertheless, it can reach $5.1\times 10^{-19}$ at an averaging time of $216,000$~s.

\section{Systematic evaluation}\label{sec:Evaluation}
\subsection{BBR Stark shift}\label{subsec:BBR}
\begin{figure}[hbtp]
\centering
\includegraphics[width=0.99\columnwidth,angle=0]{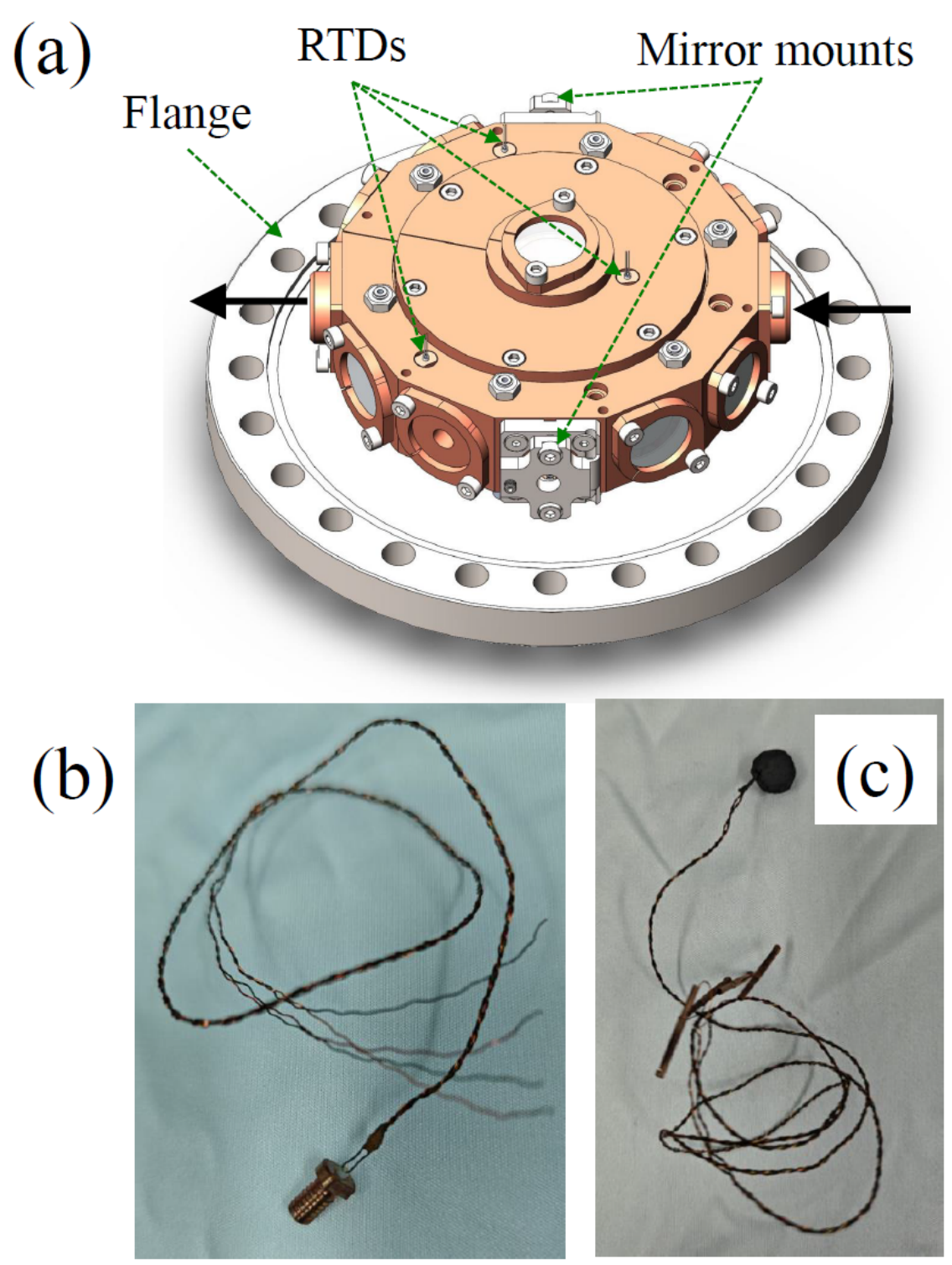}
\caption{(a) Solidworks rendering of the in-vacuum BBR shield. The black arrows represent the direction of the atomic beam. For scale, the flange is 6.75 inches in diameter. (b) Image of a platinum RTD embedded in a copper bolt that can be screwed into the copper body of the BBR shield. (c) Image of the black-coated copper sphere with a platinum RTD inside. This thermal probe was placed at the center of the BBR shield when we performed the in situ measurement of the BBR temperature.}
\label{Fig:BBR-shield}
\end{figure}
As shown in figure~\ref{Fig:BBR-shield}, the design of the in-vacuum BBR shield is similar to the previous generation~\cite{ZhangAng(APM)2022}. The internal surfaces
of the copper body are coated with nano graphite powder blended with ultra-high vacuum epoxy, while all the inside surfaces of the optical windows have been coated with indium tin oxide (ITO). Six platinum resistance temperature detectors (RTDs) are distributed sparsely on the copper body to monitor the temperature inhomogeneity of the shield. Two mirror mounts used to hold the buildup cavity mirrors have been installed on the top and bottom sides.

During vacuum baking, the RTDs in the shield have to experience a high temperature up to \mbox{130 $^{\circ}$C}. This batch of 15 RTDs (Sensing Devices; PT100/8AX*) has shown noticeable changes in resistance values after baking, corresponding to temperature reading errors up to 100~mK. Fortunately, we found that their resistance values become more stable after a repeated baking. After each baking cycle, the baking-induced resistance shift was determined by comparing their resistance with that of an un-baked reference RTD at a stabilized temperature near \mbox{23 $^{\circ}$C}. The measurement stability of these RTDs is better than 10~mK after 4 thermal cycles up to \mbox{150 $^{\circ}$C}. After the last baking cycle, the sensors were installed in the vacuum system and experienced the final bakeout of the entire vacuum system. In table~\ref{Tab:Teff-Budget} we add an additional uncertainty of 10~mK to account for the baking effect.

\begin{figure}[hbtp]
\centering
\includegraphics[width=0.99\columnwidth,angle=0]{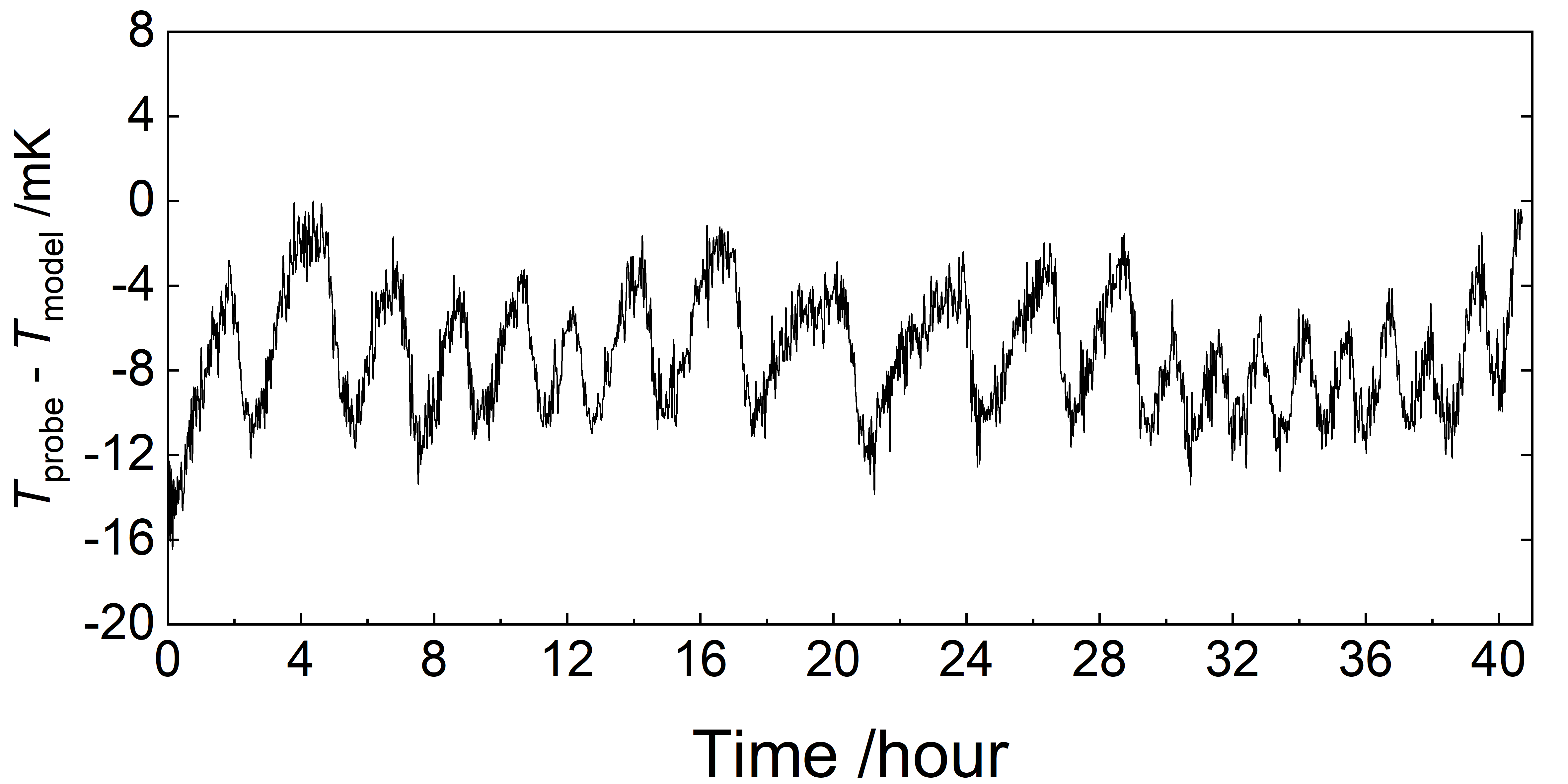}
\caption{Temporal variation of the temperature difference between the spherical probe and the BBR shield. The spherical probe is placed at the position of the atoms. Statistical averaging of these data yields $-7.1\,\mathrm{mK}$ with a standard error of 2.8~mK.}
\label{Fig:Tprob-Tmodel}
\end{figure}

The BBR shield, including optical windows on it, subtends more than 99$\%$ of the entire solid angle around the atoms, and the copper body effectively reduces the temperature inhomogeneity. Therefore, the BBR temperature felt by the atoms can be well approximated by a model temperature, $T_{\mathrm{model}}$, which is simply the average of temperature readings from the six RTDs. To check the accuracy level of the model prediction, we have carried out an in situ measurement of the BBR temperature $T_{\mathrm{BBR}}$. A calibrated RTD  embedded in a black-coated copper sphere of 8~mm in diameter was used for radiation thermometry (see figure~\ref{Fig:BBR-shield}(c)). This thermal probe was fixed on the tip of a  thin rod (3~mm in diameter) made of polyether ether ketone (PEEK). In order to produce a thermal environment as close as possible to the real vacuum system of the clock, the BBR shield was installed in a clock science chamber which was continuously pumped by a turbo pump, but without bakeout. The PEEK rod mounted on a feedthrough on the science chamber stuck into the BBR shield, with no mechanical contact to the shield's internal surfaces. The thermal probe was located at the center of the BBR shield, where atoms should reside. Since the thermal probe's disturbance to the BBR environment is negligible, $T_{\mathrm{probe}}$  can be regarded as a direct measurement of  $T_{\mathrm{BBR}}$. We monitored $T_{\mathrm{BBR}}$ and $T_{\mathrm{model}}$ for more than 40 hours, and found that $T_{\mathrm{BBR}}$ is closely followed by $T_{\mathrm{model}}$ as shown in figure~\ref{Fig:Tprob-Tmodel}. The residual difference exhibits a periodic variation that is closely correlated to the simultaneously recorded room-temperature fluctuation. Statistical averaging of the temperature differences gives a value of $-7.1(2.8)\,\mathrm{mK}$, which represents a correction that should be applied to $T_{\mathrm{model}}$. After this measurement, the thermal probe was removed, and then the science chamber, including the shield inside, was installed on the vacuum system of Yb1.

In clock operation there exist more local heat sources on the BBR shield that have not been described by the model above, i.e. the heating of buildup cavity mirrors and optical windows due to absorption of the incident laser lights. The total absorption of the two cavity mirrors to the input $759\,\mathrm{nm}$ lattice light was measured to be $\sim 12\%$, with the assumption that the scattering loss can be neglected.  We have performed finite element (FE) analysis to the mirrors, considering both the heat conductivity and thermal radiation effects. The input lattice laser power is assumed to be 156~mW, which is the maximum input power we have ever used in the experiment. The simulation shows that the temperature increases by \mbox{11 $^{\circ}$C} at the center of the beam spot on each mirror. The solid-angle-weighted variation of the BBR temperature due to the two hot spots is about 1.1~mK. Since the average input power in a clock cycle is certainly lower than the maximum value, we place an uncertainty of 1~mK on the BBR temperature to account for this heating effect. Similar FE analysis has also been applied to the optical windows where the heating effect is mainly from the absorption of the 399~nm laser light. The BBR temperature is shifted up by 0.6~mK. We conservatively place an uncertainty of 1~mK for it in table~\ref{Tab:Teff-Budget}.

\begin{table}[hbtp!]
\setlength{\tabcolsep}{1pt}
\footnotesize
\caption{BBR temperature uncertainty  for our Yb lattice clock in normal operation ($297$~K).}
\label{Tab:Teff-Budget}
\centering
%\begin{ruledtabular}
%\begin{spacing}{1.2}
\begin{tabular}{lcc}
\hline\hline
        Contributor        &  Correction/mK   &  Uncertainty/mK \\
\hline
 RTDs temperature measurements & & \\
\ \ \ \ Manufacturer calibration           & --     &   15   \\
\ \ \ \ Digital multimeter (4-wire)        & --     &   5    \\
\ \ \ \ Self heating                       & --     &   1    \\
\ \ \ \ Baking effect                      & --     &   10   \\
Model prediction &  & \\
\ \ \ \ Model error                        & -7.1   &  2.8   \\
Additional heat sources &  & \\
\ \ \ \ Lattice mirrors ($\times$2)        & --     &  1     \\
\ \ \ \ ITO-coated windows ($\times$6)     & --     &  1     \\
\hline
\textbf{Total}            & \textbf{-7.1}  &   \textbf{19}   \\
\hline\hline
\end{tabular}
%\end{spacing}
%\end{ruledtabular}
\end{table}

Under normal clock operation, the maximum difference among the readings of the six RTDs on the BBR shield is about $40$--$50\,\mathrm{mK}$, roughly consistent with the inhomogeneity of $30\,\mathrm{mK}$ observed in the in situ measurement. A comparison between normal operation and the in situ measurement shows that the relative reading of each RTD with respect to the mean value differs by less than $10\,\mathrm{mK}$, consistent with the baking-induced uncertainty of $10\,\mathrm{mK}$ assigned in table \ref{Tab:Teff-Budget}.

For a reference temperature of $T_{0}=300\,\mathrm{K}$, the BBR static shift of the Yb clock transition is $\nu_{\mathrm{stat}}=-1.254870(24)\,\mathrm{Hz}$, according to the differential static polarizability that has been measured to high accuracy by the NIST group~\cite{Sherman(NIST)2012}. Including the dynamic contribution, the BBR shift can be expressed as~\cite{Hassan(NISTcryo)2025}
\begin{equation}
\label{eq:DeltaBBR}
\Delta \nu_{\mathrm{BBR}}\left( t \right)
= \nu_{\mathrm{stat}}t^{4} + \nu_{\mathrm{dyn},6}t^{6}  +  \nu_{\mathrm{dyn},8}t^{8} + \mathcal{O}\left(t^{10}\right),
\end{equation}
where $t=T/T_{0}$, and $\nu_{\mathrm{dyn},6}=-22.17(34)\,\mathrm{mHz}$ \cite{Hassan(NISTcryo)2025}. The 8th order dynamic coefficient $\nu_{\mathrm{dyn},8}=\nu_{\mathrm{stat}}\alpha_{8}$, where $\alpha_{8}$ was measured to be 0.000593(16) \cite{Beloy(NIST)2014}. For a typical BBR temperature of $297\,\mathrm{K}$ with an uncertainty of $19\,\mathrm{mK}$ as in our case, the BBR shift is estimated to be $-1226.98(45)\,\mathrm{mHz}$, corresponding to a fractional frequency shift $-2367.34(87)\times 10^{-18}$. The total uncertainty of $8.7\times 10^{-19}$ is actually a combination of two contributions, an uncertainty of $6.1\times 10^{-19}$ in the BBR environment and an uncertainty of $6.2\times 10^{-19}$ in the atomic response.

\subsection{Density shift}\label{subsec:DensityShift}
For typical atomic preparation procedures, both the axial and the transverse temperatures of the lattice-trapped Yb atoms are at least an order of magnitude lower than the $p$-wave barrier, which was predicted to be $\sim\,30\,\mathrm{\mu K}$~\cite{Dzuba(UNSW)2010}. $p$-wave collisions are thus highly suppressed. For spin-polarized atoms with a polarization purity exceeding 99.5\% as in our case, $s$-wave collisions are also highly suppressed by the Pauli exclusion principle. However, inhomogeneous probe excitation may lead to residual $s$-wave collision shifts. To mitigate this effect, we use a probe beam with a large waist radius which is 5 times that of the lattice beam.

\begin{figure}[hbtp]
\centering
\includegraphics[width=0.99\columnwidth,angle=0]{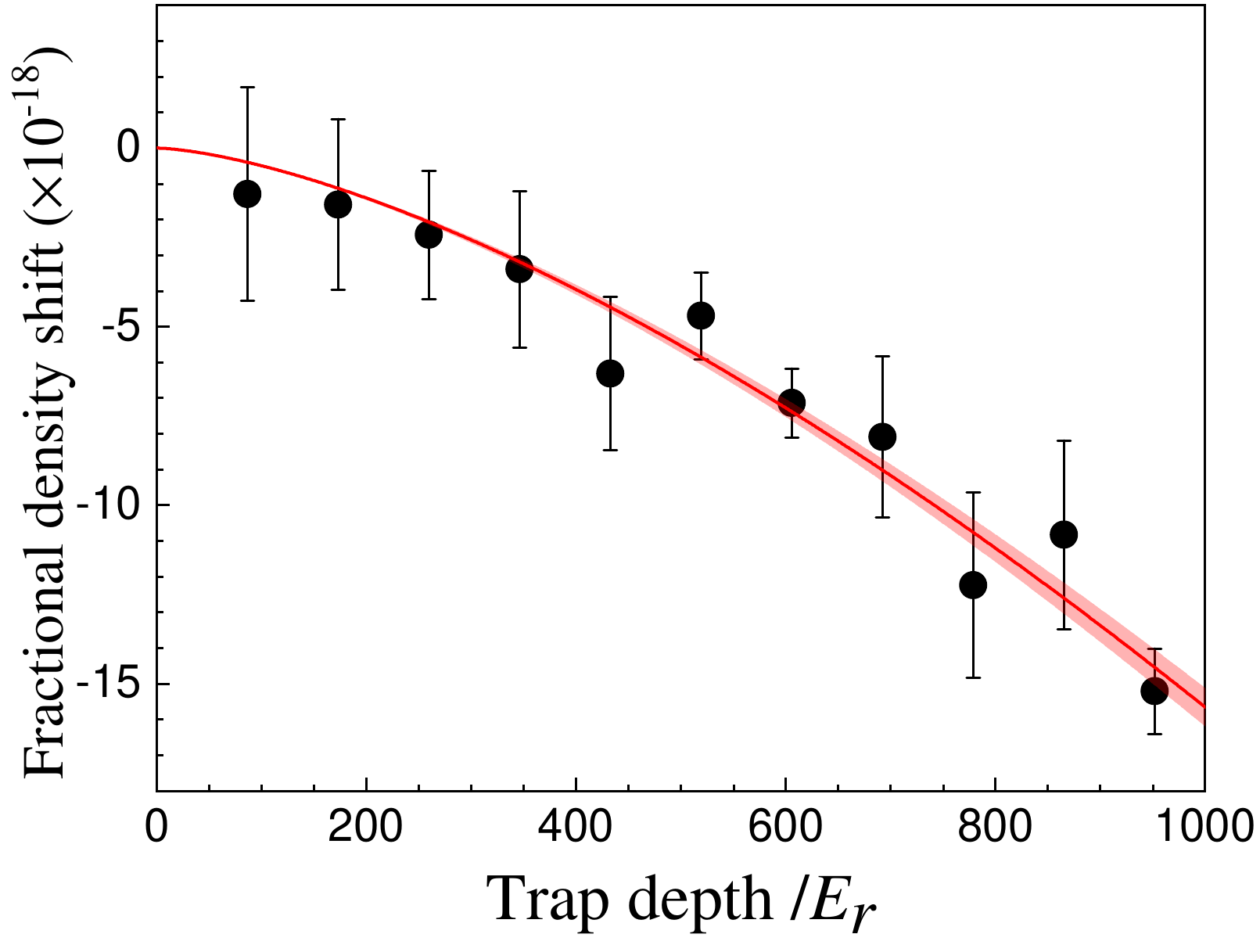}
\caption{Fractional density shift per $N_0$ atoms as a function of the lattice trap depth (black dots). These data are fit to  $\kappa U^{3/2}$ (red solid line), yielding a scaling factor $\kappa=-4.95(16)\times 10^{-22}$. The shaded area indicates the $1\sigma$ confidence band on the fit.}
\label{Fig:DensityShift}
\end{figure}

We carried out density shift measurements in Yb2 for different trap depths, with the atom number interleaved between the low ($0.5 N_0$) and high ($5 N_0$) cases. Here, $N_0$ is the reference atom number corresponding to a specific reading from the photo multiple tube (PMT) used to detect the atoms. We have calibrated the atom number based on the 399~nm fluorescence signal, and determined that $N_0\approx11,000$. Black dots in figure~\ref{Fig:DensityShift} denote the measured fractional clock shift per $N_0$ atoms, and the error bars are determined from the last point of the total Allan deviation of each self-interleaved measurement.

The measured density shifts that have been normalized to $N_0$ atoms in figure \ref{Fig:DensityShift} are fitted to $\kappa U^{3/2}$ (red solid line), giving a scaling factor of $\kappa=-4.95(16)\times 10^{-22}$ and a reduced chi-squared $\chi^2_{\mathrm{red}}=0.32$, with $U$ the dimensionless ratio of trap depth to the recoil energy $E_r$. The fractional density shift $\delta \nu_{\mathrm{col}}/\nu_{\mathrm{clock}}$ can then be expressed as $(N / N_0)\kappa U^{3/2}$. Under the normal operation conditions, $N = 1.00(15)N_0$ and $U=52(1)$, the density shift is $-1.9(3)\times 10^{-19}$.

\subsection{Lattice light shift}\label{subsec:LatticeLight}
Beloy \textit{et al.} developed a nonperturbative model, the so-called BO + WKB model~\cite{Beloy(NIST)2020}, to accurately determine the motional energy spectra of atoms trapped in a one-dimensional optical lattice. They used this model to characterize lattice light shifts in OLCs, considering the electric dipole ($E1$) coupling, magnetic dipole ($M1$) and electric quadrupole ($E2$) coupling, as well as the hyperpolarizability coupling. They have derived an expression for the lattice light shift $\Delta\nu_{\mathrm{lat}}$ in the following form:
\begin{eqnarray}
\label{eq:LatticeShift-XYZ}
\frac{h\Delta\nu_{\mathrm{lat}}}{E_r}= & -U\frac{1}{\alpha_{E1}}\frac{\partial\Delta\alpha_{E1}}{\partial\nu_l}\left(\nu_l-\nu_{E1}\right) X\nonumber\\
& - U\frac{\Delta\alpha_{qm}}{\alpha_{E1}} Y - U^{2}\frac{\Delta\beta E_{r}}{\alpha^{2}_{E1}}Z,
\end{eqnarray}
where $\alpha_{E1}$ is the $E1$ polarizability, and $\nu_{E1}$ represents the magic frequency, while $\Delta\alpha_{E1}$, $\Delta\alpha_{qm}$ and $\Delta\beta$ are differential atomic parameters between the clock states, corresponding to the $E1$ polarizability, the sum of $M1$ and $E2$ polarizability, and the hyperpolarizability, respectively. The three factors $X$, $Y$, and $Z$ describe the external degree of freedom of the trapped atoms, and they all depend on both the axial temperature $T_{z}$ and the radial temperature $T_{\rho}$. As long as the relation between the temperature and the trap depth $U$ is given, $X$, $Y$, and $Z$ can all be numerically calculated.

Under the conditions that both $T_{z}$ and $T_{\rho}$ scale linearly with $U$, factors $X$, $Y$, and $Z$ are approximately constant over a large depth range. Then the expression of $\Delta\nu_{\mathrm{lat}}$ can be reparameterized into a simple form~\cite{Brown(NIST)2017}
\begin{equation}
\label{eq:TwoPara-LatticeShift}
\frac{\Delta\nu_{\mathrm{lat}}}{\nu_{\mathrm{clock}}} = -\alpha^{*}U - \beta^{*}U^{2},
\end{equation}
where $\alpha^{*}$ and $\beta^{*}$ are the linear and quadratic parameters, respectively. This two-parameter model had been used to analyze the lattice light shifts in previous works. However, the required conditions are not satisfied in our clocks. We have measured the radial temperature, and found that $T_{\rho}\propto \sqrt{U}$. This relationship for $T_{\rho}$ versus $U$ is consistent with the adiabatic ramp of lattice intensity. Motional sideband cooling along the axial dimension pumps nearly all atoms to the ground state of $n_{z}=0$, and the fractional populations on  each $n_{z}$ state keep constant during the subsequent adiabatic ramp process.

For $T_{\rho}$ scaling as $\sqrt{U}$, factors $X$, $Y$, and $Z$ are not constant anymore. In this case equation (\ref{eq:TwoPara-LatticeShift}) would lead to a model error at the $10^{-18}$ level when used to fit the measured data. However, as suggested by Beloy \textit{et al.}, the error can be reduced by increasing the number of fitting parameters~\cite{Beloy(NIST)2020}. To account for the dependence of the three factors on the lattice depth, we assume that each factor can be expanded to the second order in $U$. Then, $\Delta\nu_{\mathrm{lat}}$ can be described by a fourth-order polynomial with four fitting parameters
\begin{eqnarray}
\label{eq:LatticeShift-Fractional}
\frac{\Delta\nu_{\mathrm{lat}}}{\nu_{\mathrm{clock}}}= -\alpha_{1}U - \alpha_{2}U^{2}
  - \alpha_{3}U^{3} - \alpha_{4}U^{4},
\end{eqnarray}
where the first three parameters $\alpha_{1}$, $\alpha_{2}$, and $\alpha_{3}$ are linear functions of $\nu_l$. At a specific frequency $\nu_\mathrm{zero}$, which is close to the magic frequency $\nu_{E1}$, $\alpha_{1}$ can eliminate the linear term of the lattice light shift, as described by
\begin{equation}
\label{eq:alpha1}
\alpha_1(\nu_l)=\frac{\partial\alpha_1}{\partial\nu_l}\left( \nu_{l}-\nu_{\mathrm{zero}}\right).
\end{equation}

\begin{figure}[hbtp]
\centering
\includegraphics[width=0.99\columnwidth,angle=0]{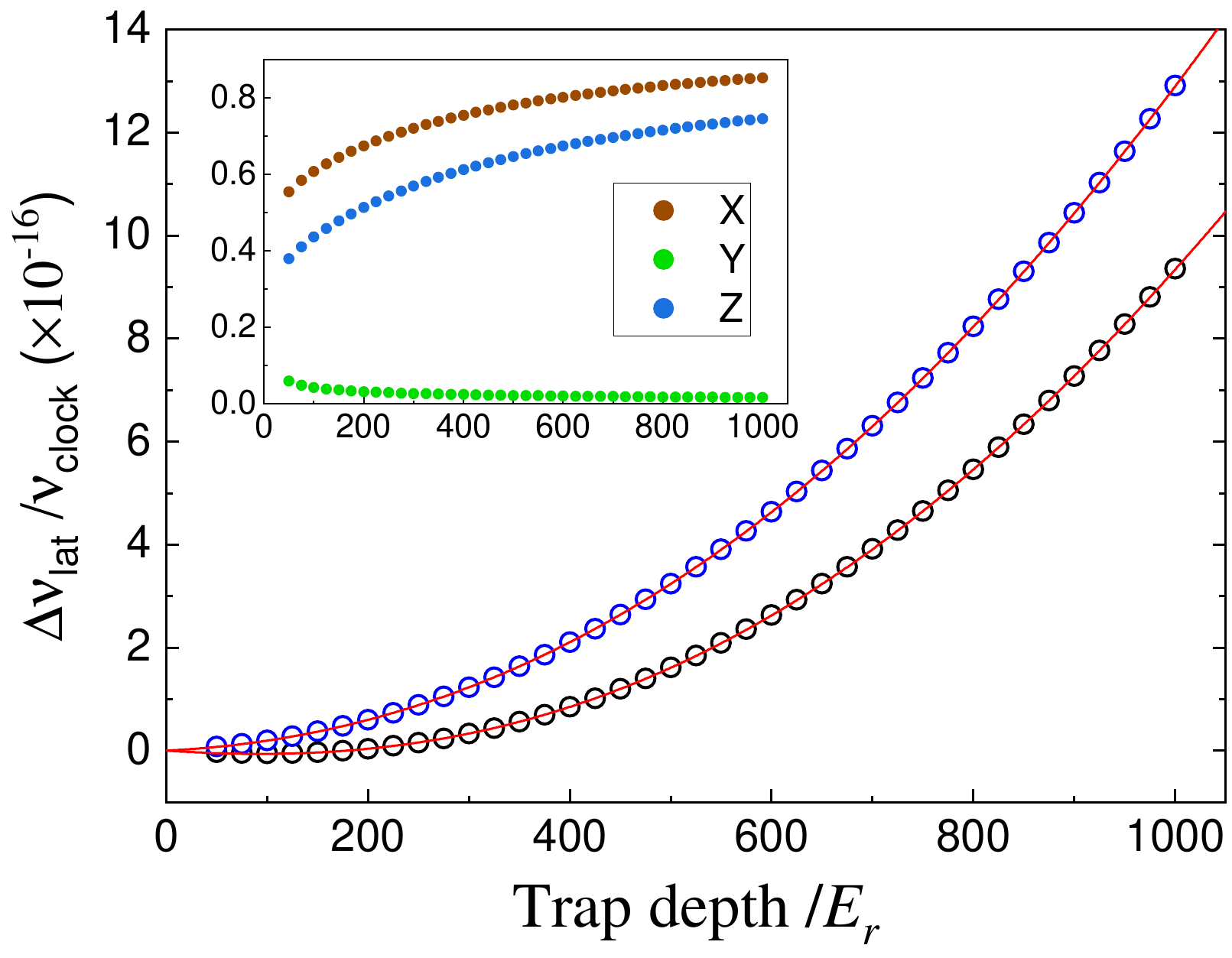}
\caption{Lattice light shifts (open circles) versus the trap depth, numerically calculated using the BO + WKB model for a radial temperature of $T_{\rho}=\sqrt{U}/3$~($\mu$K). The lattice light is assumed to be detuned from $\nu_{E1}$ by 3~MHz (black open circles) or $-3$~MHz (blue open circles). Fractional populations of the axial motional states are assumed to be $\langle n_0\rangle=0.96$ and $\langle n_1\rangle=0.04$. Each of the two datasets is fitted with a fourth-order polynomial in the form of equation~(\ref{eq:LatticeShift-Fractional}) (the red curves). The inset displays the calculated values of $X$, $Y$ and $Z$.}
\label{Fig:BO-WKB_LatticeShift}
\end{figure}

To check the model error of equation (\ref{eq:LatticeShift-Fractional}), we first generate the frequency shift data from the BO + WKB model. The data set is intended to simulate experimentally obtained data. In the calculation of $\Delta\nu_{\mathrm{lat}}$, we have adopted the values of $\alpha_{E1}$, $\alpha_{qm}$, $\beta$ and $\partial\Delta\alpha_{E1}/\partial\nu_l$ used in~\cite{Katori(RIKEN)2015}. The radial temperature $T_{\rho}$ is assumed to scale as $\sqrt{U}/3$ (in units of $\mu$K), which basically matches the measured temperature data. The data set of $\Delta\nu_{\mathrm{lat}}$ in the depth range of 50 to 1000~$E_r$ is then fitted with equation (\ref{eq:LatticeShift-Fractional}). The deviation of the fitting curve from the data set indicates the accuracy level at which the four-parameter model can reproduce the true frequency shifts.  Two typical data sets and the corresponding fitting curves are displayed in figure~\ref{Fig:BO-WKB_LatticeShift}. Analysis of a series of data sets shows that the four-parameter model supports accuracy at the $10^{-19}$ level.

\begin{figure}[hbtp]
\centering
\includegraphics[width=0.99\columnwidth,angle=0]{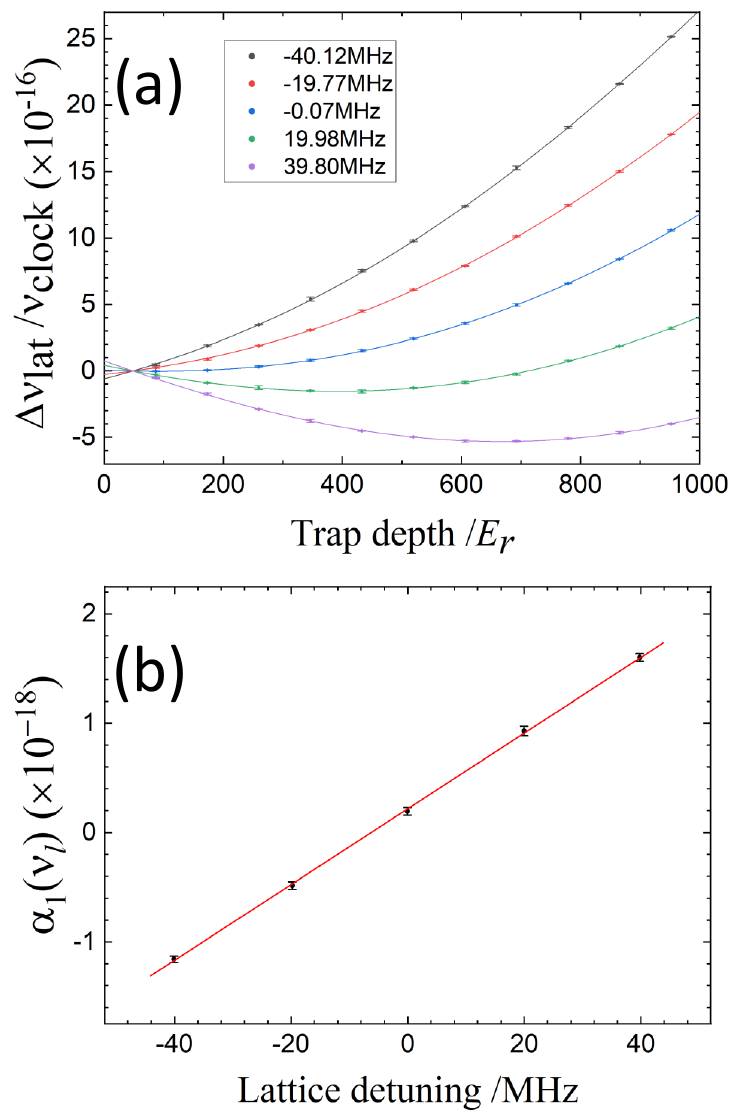}
\caption{(a) Lattice light shifts as a function of lattice depth. Colored dots represent data sets with distinct detunings of $\nu_{l}$ relative to the reference frequency $394~798~266.9~\mathrm{MHz}$. Solid lines represent global fits with equation (\ref{eq:LatticeShift-Fractional}). (b) Coefficient $\alpha_1$ from the global fit as a function of the detuning of $\nu_{l}$. The solid line is a linear fit.}
\label{Fig:LatticeShift}
\end{figure}

We measured the lattice light shift of Yb2 at varied depth $U$, by using the clock frequency of Yb1 as a reference. The lattice depth of Yb1 was set to a fixed value of $48\, E_r$, while the depth of Yb2 varied from  $87\, E_r$ to $952\, E_r$. The two clocks shared the same lattice laser. The detuning of $\nu_l$ is defined relative to $394~798~266.9~\mathrm{MHz}$, the value of $\nu_{E1}$ determined by the NIST group~\cite{Bothwell(NIST)2025}. In the configuration of synchronous comparison, the differential frequency between Yb2 and Yb1 can reach a resolution of the order of $10^{-18}$ in a moderate measurement time. Density shift has been subtracted for each data point. Five datasets were taken for distinct detunings: $-40.12$, $-19.77$, $-0.07$, $19.98$, and $39.80~\mathrm{MHz}$, as shown in figure \ref{Fig:LatticeShift}(a).

We have performed a global fit to the dataset using equation~(\ref{eq:LatticeShift-Fractional}), considering also the frequency shift of Yb1 at different detunings. The coefficients $\alpha_{1}$, $\alpha_{2}$, and $\alpha_{3}$ were fitted independently for each dataset, while $\alpha_{4}$ was taken as a global parameter. The solid lines in figure~\ref{Fig:LatticeShift}(a) are the fitting curves that match the datasets. Five values of $\alpha_{1}$ from the fit have been displayed in figure~\ref{Fig:LatticeShift}(b) as a function of the detuning. The linear fit to the five data points yields a slope of $\partial\alpha_{1}/\partial\nu_l=3.46(4)\times 10^{-20}/ \mathrm{MHz}$, as well as a zero-linear-shift frequency $\nu_{\mathrm{zero}}=394~798~260.6(3)~\mathrm{MHz}$. This value of $\nu_{\mathrm{zero}}$ is within $1~\mathrm{MHz}$ of $\nu_{E1}$ as determined by the RIKEN group \cite{Nemitz(RIKEN)2019}. As shown in table~\ref{tab:LatticeShift}, operating at $\nu_{\mathrm{zero}}$ and a lattice depth of $52(1)\,E_r$, the lattice light shift of Yb2 is calculated to be $3.04\times 10^{-18}$, with an uncertainty of $6.3\times 10^{-19}$ which is mainly limited by the uncertainty in $\nu_{\mathrm{zero}}$.

\begin{table}[htbp!]
\setlength{\belowcaptionskip}{0.05cm}
\setlength{\tabcolsep}{3pt}
%\\renewcommand{\arraystretch}{1.1}
\caption{Lattice light shift uncertainty budget for Yb2 under typical operating condition.}
%\small
\label{tab:LatticeShift}
\footnotesize
\begin{tabular}{lccc}
\hline \hline
Items                                  & Value          & Error   & Uncertainty ($10^{-19}$)\\ \hline
$\frac{\partial\alpha_{1}}{\partial\nu_l} (10^{-20}/ \mathrm{MHz})$
                                       & 3.46           & 0.04    & 0    \\
$\nu_{\mathrm{zero}}\ (\mathrm{MHz})$  & 394~798~260.6  & 0.3     & 5.4  \\
$\alpha_2\ (10^{-21})$                 & -1.10          & 0.11    & 2.9  \\
$\alpha_3\ (10^{-25})$                 & -5.6           & 0.8     & 0.1  \\
$\alpha_4\ (10^{-28})$                 & 2.4            & 1.0     & \textless 0.01  \\
$U$                                    & 52             & 1       & 1.2  \\
$\nu_{l}\ (\mathrm{MHz})$              & 394~798~260.6  & 0.05    & 0.9  \\
  \hline
  \textbf{Shift} $(10^{-19})$          & \textbf{30.4}  &         & \textbf{6.3}\\
 \hline \hline
\end{tabular}
\par\smallskip
The error of $\partial\alpha_{1}/\partial\nu_l$ has no contribution when $\nu_{l}=\nu_{\mathrm{zero}}$. The values and errors of parameters $\alpha_2$ and $\alpha_3$ are evaluated at $\nu_{\mathrm{zero}}$. The error in $\nu_{l}$ originates from the accumulated drift in the cavity resonance frequency to which the lattice laser is locked, over a typical clock-operation period of several days.
\end{table}

\subsection{Other systematic effects}\label{subsec:OtherSystematics}
\noindent \textbf{Zeeman shift.} We had measured both the linear and the quadratic Zeeman shift (QZS) coefficients in our previous system \cite{Zhangang(APM)2023}, and the obtained values are consistent with those reported by the NIST group~\cite{McGrew(NIST)2018}. We use a magnetic field of 1~G to define the quantization axis for our new clocks. By alternating interrogations between the opposite spin states, the first-order Zeeman shift is cancelled completely for a DC magnetic field. However, slow drift of the magnetic field leads to a clock frequency shift through the servo loop. The observed frequency splitting shows a varied drift rate during a day, with a maximum value of about $10\,\mu\mathrm{Hz/s}$. We thus place an upper bound of $1\times 10^{-19}$ on the residual Zeeman shift. Our magnetic field can be determined on an accuracy level of 0.3~mG. According to the recently updated QZS coefficient ($-0.05997(7)\,\mathrm{Hz/G^{2}}$)~\cite{Bothwell(NIST)2025}, QZS at 1~G corresponds to a shift of $-115.71(15)\times 10^{-18}$. During normal clock operation, the QZS is corrected in real time.

\noindent \textbf{Lattice vector.} Vector Stark shift from the lattice laser that acts as a pseudo-magnetic field is highly suppressed due to the linear polarization of the lattice light. Since we have applied a power stabilization to the intra-cavity lattice beam, the residual vector shift is constant, and thus can be cancelled completely, just as a DC magnetic field. However, this small shift affects the precise determination of the magnetic field, leading to an extra uncertainty on the QZS. We measured the residual vector shift with Yb2.
By self-interleaving two lattice depths of $52\,E_{r}$ and $952\,E_{r}$, the differential Zeeman splitting was measured, giving a vector shift coefficient of 0.28(6)~$\mathrm{mHz}/E_{r}$. At the typical lattice depth of $52\,E_{r}$, the residual vector shift is only 15(3)~$\mathrm{mHz}$, contributing an uncertainty of $1.7(4)\times 10^{-20}$ to the QZS.

\noindent \textbf{Doppler shift.} 578~nm clock laser light is delivered to the experimental table through an optical fiber. The reference mirror used for fiber-noise cancellation (FNC) is not very close to the buildup cavity, having an optical length of about 80~cm in between. Vibrations of the cavity mirrors are highly suppressed by the experiment table. Based on the estimation of thermal expansion induced variation in optical length, we place an uncertainty of $6\times 10^{-19}$ for the first-order Doppler shift.

\noindent \textbf{Background gas.} Background gases (dominated by $\mathrm{H}_{2}$), when colliding with the cold Yb atoms, lead to a shift of the clock frequency. This shift was confirmed to scale with the trap loss rate, $\Gamma$, with a coefficient of $-1.64(12)\times 10^{-17}\,\mathrm{s}$~\cite{McGrew(NIST)2018}. For Yb2, we repeated the trap lifetime measurement for 30 times. The obtained average value of $1/\Gamma=10.5(3)\,\mathrm{s}$ corresponds to a shift of $-1.56(12)\times 10^{-18}$.

\noindent \textbf{DC Stark shift.} Our BBR shield has been coated with conductive graphite on its internal surfaces, while each optical window on the shield has a conductive ITO coating on the internal surface. They form an almost fully enclosed Faraday cage to shield stray electric fields. The design of BBR shield was first developed by the NIST group, and their measurements confirmed that the Stark shift due to stray electric field can be constrained to the order of  $10^{-20}$~\cite{Beloy(NIST)2018}.  We thus put an upper bound of $1\times 10^{-19}$ on the DC Stark shift.

\noindent \textbf{Probe Stark shift.} For Rabi spectroscopy, the probe AC Stark shift is inversely proportional to the square of the interrogation time. Since 560-ms Rabi spectroscopy corresponds to a probe shift of $2\times 10^{-20}$ \cite{McGrew(NIST)2018}, our longer interrogation time of 800~ms should correspond to a shift of $\sim 1\times 10^{-20}$. The uncertainty level can be conservatively assessed at $50\%$ of this value.

\noindent \textbf{Line pulling.} Spin state preparation can be performed with a high purity better than $99.5\%$. Line pulling effects of the undesired spin states are negligible due to the small residual population and large detunings from the desired transition. The closest transverse sidebands are detuned by only $30\,\mathrm{Hz}$ from the carrier. However, their spectroscopic features, as pointed out in Ref~\cite{McGrew(NIST)2018}, are highly symmetric and largely suppressed due to the collinear propagation of the probe and lattice beams. Therefore, an upper bound of $1\times 10^{-19}$ can be placed on the uncertainty from line pulling.

\noindent \textbf{Servo error.} Local oscillator frequency drift is cancelled to a level well below $1\,\mu\mathrm{Hz/s}$ by digital feed-forward correction. Analysis of the lock error signal during long-term operation indicates that servo error is consistent with zero at a level of $-1(4)\times 10^{-20}$.

\noindent \textbf{AOM chirp.} The AOM used to generate the probe pulse is operated at a low RF power of tens mW. Outside the interrogation time, the RF input is detuned by several MHz, rather than being switched off. The low and constant RF power prevents the probe beam from thermal-transient induced phase chirp. In addition, since the probe beam derived from the ‑1st-order of the AOM is outside the FNC servo loop (which uses the 0th order), no noticeable phase oscillation was observed during frequency switching at the beginning of interrogation. We thus use a conservative upper bound of $1\times 10^{-19}$ to account for any residual phase chirp.

\noindent \textbf{Background light shift.} Each of the two 759~nm lattice lasers is filtered by a volume Bragg grating (VBG) bandpass filter (OptiGrate; SPC-759) which has a bandwidth of 21~GHz and a peak reflectivity of $96\%$. Similar VBG had been applied to tapered amplifier (TA) laser systems, with the background light shift suppressed well below $1\times10^{-18}$~\cite{Fasano(NIST)2021,Jia(USTC)2025}. Compared to a TA laser system, which represents a spectrally dirty case, our fiber-laser-based lattice light source is intrinsically purer. In addition, for cavity enhanced lattice light, the background spectrum is further suppressed by a factor comparable to the enhancement factor. We can thus assign a conservative upper bound of $1\times 10^{-19}$ to the background light shift.
\begin{table}[htbp!]
\setlength{\belowcaptionskip}{0.05cm}
\caption{Uncertainty budget of our $^{171}$Yb clock (Yb2) for typical operating conditions. Values marked with * are corrected in real time.}
\label{tab:YbErrorBudget}
%\small
\footnotesize
\begin{tabular}{lcc}
\hline \hline
Effect & Shift ($10^{-18}$)  & Uncertainty ($10^{-18}$)  \\ \hline
  BBR*              & -2367.3    &     0.87       \\
  Density           & -0.19      &     0.03       \\
  Lattice light     & 3.04       &     0.63        \\
  1st order Zeeman  & 0          &   \textless 0.1       \\
  2nd order Zeeman* & -115.7     &     0.15       \\
  Lattice vector    &  -0.017    &     0.004      \\
  1st order Doppler &  0         &     0.6        \\
  Background gas    & -1.56       &     0.12       \\
  DC Stark          &     0      &   \textless 0.1  \\
  Probe Stark       &  0.01      &   0.005          \\
  Line pulling      &   0        &   \textless 0.1   \\
  Servo error       &  -0.01     &  0.04          \\
  AOM chirp         &  0         &   \textless 0.1 \\
  Background light  &  0       &    \textless 0.1           \\
  \hline
  \textbf{Total} & \textbf{-2481.7} & \textbf{1.3}\\
 \hline \hline
\end{tabular}
\end{table}

With all significant systematics evaluated, we present the uncertainty budget of Yb2 in Table~\ref{tab:YbErrorBudget}. The total systematic uncertainty of $1.3\times 10^{-18}$ meets the $2\times 10^{-18}$ threshold set for redefinition of the second. Among the systematic effects listed, only the BBR shift and the quadratic Zeeman shift are corrected in real time. Careful control of these two systematics is critical for achieving high frequency stability at long timescales.

\section{Discussion and summary}
We have described the systematic evaluation of our newly built $^{171}$Yb lattice clock with a total systematic uncertainty of $1.3\times 10^{-18}$. The high frequency stability allows quick measurement of the systematic shifts. Magic frequency of the optical lattice has been determined at an accuracy level of 0.3~MHz. The lattice light shift, which often represents the leading clock systematic error, has been precisely measured, reaching an uncertainty below $10^{-18}$. This is achieved by cavity-enhanced optical lattice and the differential frequency measurement between two clocks. Density shift of the trapped atoms is reduced due to the larger trap volume of the optical lattice. Our clocks will be used for remote frequency comparison with a laboratory-based $^{87}$Sr lattice clock in Shanghai via a newly established telecom fiber link.

\section*{Acknowledgments}
We are grateful to Han-Ning Dai and Hai-Feng Jiang for valuable discussions on the evaluation of systematic uncertainty. Special thanks to Yan-Yi Jiang for lending the RTDs. This work is supported by the Strategic Priority Research Program of the Chinese Academy of Sciences under Grant No. XDA0250102, by the National Science and Technology Major Project under Grant No. 2025ZD0300702, by the Natural Science Foundation of China under Grant No. 12374467, and by Quantum Science and Technology-National Science and Technology Major Project (Grant No. 2021ZD0300902).

\section*{References}
\bibliography{References_Revised}

@article{Ido2016,
   author = {T. Ido and H. Hachisu and F. Nakagawa and Y. Hanado},
   title = {Rapid evaluation of time scale using an optical clock},
   journal = {Journal of Physics: Conference Series},
   volume = {723},
   pages = {012041},
   DOI = {10.1088/1742-6596/723/1/012041},
   year = {2016},
   type = {Journal Article}
}

\end{document}